\documentclass[aps,prb,twocolumn,superscriptaddress,amsmath,amssymb]{revtex4-2}
\usepackage{booktabs}
\usepackage{graphicx}
\usepackage{gensymb}
\usepackage{tabularx}

\usepackage{nicefrac}
\usepackage{hyperref}
\hypersetup{colorlinks=true, breaklinks=true, citecolor=blue, linkcolor=blue, menucolor=blue, urlcolor=blue}
\usepackage{ulem}
\newcommand{\twobytwo}{(2$\times$2)}

\begin{document}

\title{When Surface Dynamics Fakes Symmetry - Oxygen on Rh(100) Revisited }

\author{Lutz Hammer}
\email{lutz.hammer@fau.de}
\affiliation{Solid State Physics, Friedrich-Alexander-Universit\"{a}t Erlangen-N\"{u}rnberg,
	91058 Erlangen, Germany}

\author{Tilman Ki{\ss}linger}
\affiliation{Solid State Physics, Friedrich-Alexander-Universit\"{a}t Erlangen-N\"{u}rnberg,
	91058 Erlangen, Germany}

\author{Margareta Wagner}
\affiliation{ Institute of Applied Physics, TU Wien, 1040 Vienna, Austria}

\author{\mbox{Reinhard B. Neder}}
\affiliation{Crystallography and Structural Physics, Friedrich-Alexander-Universit\"{a}t Erlangen-N\"{u}rnberg, 91058 Erlangen, Germany}

\author{Michael Schmid}
\affiliation{ Institute of Applied Physics, TU Wien, 1040 Vienna, Austria}

\author{Ulrike Diebold}
\affiliation{ Institute of Applied Physics, TU Wien, 1040 Vienna, Austria}

\author{M. Alexander Schneider} 
\affiliation{Solid	State Physics, Friedrich-Alexander-Universit\"{a}t Erlangen-N\"{u}rnberg,
	91058 Erlangen, Germany}

\begin{abstract}
Heating a long-range ordered adsorbate phase beyond its stability temperature does not necessarily result in a disordered phase, it can also break up into heavily fluctuating ordered domains. 
Temporal and/or spatial averaging over these fluctuations may give the impression of both a wrong periodicity and a false local symmetry. This can happen even below liquid-nitrogen temperature, so that the true nature of the phase might remain undetected. We demonstrate this scenario at the catalytically active Rh(100) surface covered by  $\nicefrac{1}{2}$  monolayer (ML) of oxygen, using quantitative low energy electron diffraction (LEED), variable-temperature scanning tunneling microscopy (STM) and density functional theory (DFT). Using the example of CO adsorption, we show that local symmetry can have a decisive influence on the binding energy and thus the chemical reactivity.
\end{abstract}

\maketitle

The key for any understanding of catalytic reactions is the determination of active sites. 
These may either exist a priori at the catalyst's surface or evolve during the reaction via surface restructuring. 
The decisive role of such reconstructions in determining surface chemistry and catalytic activity has been known for long \cite{Somorjai1995} and hence a reliable structure determination of adsorbate phases on surfaces is paramount. One might assume that nowadays most structures at least of rather simple adsorbate systems on relevant model catalyst surfaces have been conclusively determined, since powerful computers enable a quantitative structural analysis both experimentally by diffraction methods (aided by real-space STM imaging) or theoretically by efficient DFT codes. 
Today's precision of theoretical modeling is sometimes considered even superior to experiment for revealing the true surface structure \cite{Tan2018}. 
However, the predictive power of DFT is not always given and hence experimental techniques like quantitative LEED and STM are still urgently needed at least to elucidate the size and symmetry of the unit cell as well as the detailed atom positions.

In this study we demonstrate that experimental methods may come up with incorrect answers when surface dynamics mask the true structure by giving the appearance of a false periodicity and/or local symmetry. 
This may readily occur when thermal activation creates a multitude of domain boundaries within a long-range ordered phase rather than random disorder.
In such a case LEED observes just a coherent superposition of snapshots from all domains within the coherence width. STM, on the other hand, measures a temporal average over domains fluctuating at time scales that are short compared to typical STM data acquisition times. 
Both these averages do not necessarily reflect the structural properties of any single domain. Thus, also the true nature and energetics of e.g. available adsorption sites can be masked. 
We will demonstrate here that a Rh(100) surface covered by 0.5\,ML oxygen atoms represents such a system. It is very likely that more systems of that type exist that have remained undetected so far. 

\begin{figure}[tb]
	\centering
	\includegraphics[width=0.95\columnwidth]{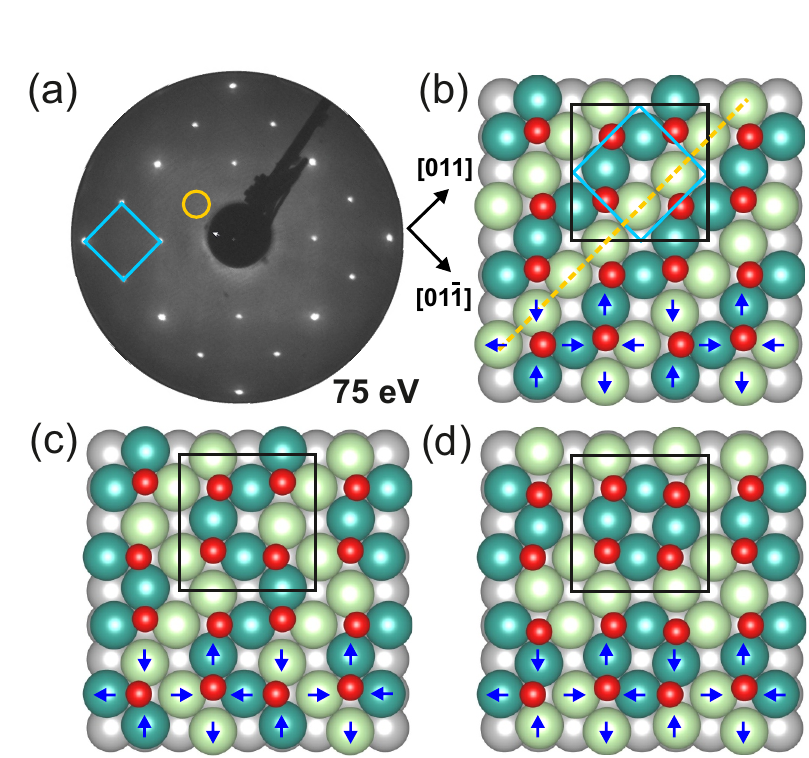}
	\caption{(a) LEED pattern of the Rh(100) surface covered with 0.5 ML oxygen taken at $100\,\textrm{K}$. 
	The pattern suggests a \twobytwo\ unit cell (blue) with two orthogonal glide planes that cause extinction of the $(m+\nicefrac{1}{2},~ 0)$ and $(0,~n+\nicefrac{1}{2})$ beams (the position of one is marked by a yellow circle). 
	(b) Top view of the \twobytwo-2O structure resulting from the LEED analysis of Ref.\,\cite{Baraldi1999}. 
	This model possesses one glide plane (yellow) only. 
	(c) and (d) Two proposed variants of the model (b) that enlarge the unit cell to a Rh(100)-$(2\sqrt{2} \times 2\sqrt{2})R45^\circ$-4O structure (black). 
	The structures are obtained from (b) by letting the lower right (c) and also the upper right (d) oxygen atoms of the ``$2\sqrt{2}$'' cell hop to an equivalent site of the reconstructed Rh(100) surface. 
	In all models (b)--(d) half of the Rh atoms in the top-layer are singly coordinated with oxygen (colored light green) and the other half are doubly coordinated (dark green). 
	The arrows in (b--d) indicate the lateral relaxation pattern of the top layer Rh atoms. 
	}
	\label{LEED_Models}
\end{figure}

Clean and adsorbate-covered Rh surfaces are popular model catalysts since Rh is catalytically active in a variety of chemical reactions, e.g. inducing NO$_x$ reduction \cite{Shelef1994} or alkane production \cite{Davda2005}. 
In carbon monoxide oxidation reactions the activity is found to switch from low to high just when the coverage of surface oxygen reaches 0.5\,ML \cite{Gustafson2013}. 
At this coverage the Rh(100) surface starts to reconstruct and forms threefold-coordinated adsorption sites, which are precursor states for surface sub-oxide and oxide phases developing at higher oxygen coverage (0.66\,ML \cite{Gustafson2012,Kisslinger2017} and 1.75\,ML \cite{Gustafson2005}). 
The oxygen-induced reconstruction has therefore raised great experimental and theoretical interest \cite{Oed1988,Mercer1996,Baraldi1997,Shen1998,Alfe1998,Baraldi1999,Alfe1999,Norris2000,Kirsch2004,Bianchettin2009,Tan2017,Tan2018}. 
This phase shows a (2$\times$2) LEED pattern with systematic extinction of half-order spots along the two k-axes [Fig.\,\ref{LEED_Models}(a)], indicating two perpendicular glide planes within the structure. 
The currently favored structural model based on the work of Alfè et al.\ (DFT \cite{Alfe1998}), Baraldi et al.\ (LEED \cite{Baraldi1999}) and Norris et al.\ (X-ray diffraction XRD \cite{Norris2000}) is depicted in Fig.\,\ref{LEED_Models}(b): The first-layer Rh atoms form the characteristic ``clock" reconstruction creating pseudo three-fold hollow adsorption sites for oxygen atoms. 
 
However, structural parameters obtained by the different methods do not coincide within the mutual error margins. Even more severe is the fact that
the model has one glide plane only and the incoherent superposition of the expected $90^\circ$ rotational domains would not lead to the extinction of the $(m+\nicefrac{1}{2},~ 0)$ and $(0,~n+\nicefrac{1}{2})$ LEED beams ($m,n \in \mathbb{Z}$) as observed in experiment. 

In the light of these discrepancies and noting the only very moderate fit quality of the LEED study mentioned ($R_\text{Pendry} = 0.28$) \cite{Baraldi1999} we decided to revisit this system in a concerted effort applying LEED intensity analysis (LEED-I(V)), variable-temperature STM and DFT. A brief description of the experimental and computational procedures is given in the appendix, more details can be found a forthcoming publication \cite{Hammer2025}.

The 0.5\,ML oxygen phase was prepared by dosing the clean Rh(100) surface with molecular oxygen either at room temperature until no further dissociation takes place (at about $9 \cdot 10^{-7}$\,mbar$\cdot$s) or, alternatively, at a temperature of 870\,K and pressure of $1 \cdot 10^{-6}$\,mbar O$_2$, which is beyond the stability limit of the next denser ($3 \times 1$)-2O phase \cite{Kisslinger2017} and thus also self-limiting. 
The latter produced an extremely well-ordered and clean surface as proven by our low-temperature STM analysis. 
The LEED data of the two preparation methods did not show any noticeable differences.  

When we tested our LEED data against the proposed \twobytwo\ model of Fig.\,\ref{LEED_Models}(b) we surprisingly arrived at an excellent Pendry R-factor of $R = 0.10$ with essentially the same model parameter values as given by Baraldi et al.\ \cite{Baraldi1999}. 
This leads us to believe that some technical error has occurred in that previous analysis. 
However, we cannot corroborate their statement that the half-order spots along the k-space axes resulting for this model structure are too weak to be detected in experiment; in contrast, their predicted intensity is clearly sufficient for observation (for details see ref.\,\cite{Hammer2025}). 
Therefore, even the excellent fit quality of the LEED analysis achieved here is not able to eliminate the discrepancies described above.

How is it possible that a structural model matches the quantitative diffraction data almost perfectly, yet is still incompatible with the symmetry observed in diffraction? The good correspondence between experimental and theoretically predicted intensities excludes significant variations of local atomic positions of both Rh and O atoms without heavily deteriorating the fit. 
The only way out is keeping the Rh clockwork reconstruction and finding new models with a different spatial arrangement of oxygen atoms among the three-fold hollow sites, which exist on either side of the long-bridge positions. 
This is in line with previous suggestions that this system may be dynamic with local disorder except at very low temperatures \cite{Alfe1999,Baraldi1999}.
Note that a new \textit{ordered} model cannot be defined in the \twobytwo\ cell since any site switch of one of the two oxygen atoms in the cell always results in a symmetry-equivalent cell. 
The next larger cell where two further nonequivalent configurations can be found is a $(2\sqrt{2} \times 2\sqrt{2})\mathrm{R}45^\circ$ cell (hereinafter referred to as ``$2\sqrt{2}$'' for short), depicted in Fig.\,\ref{LEED_Models}(c) and (d).

\begin{figure}[htb]
	\centering
	\includegraphics[width=\columnwidth]{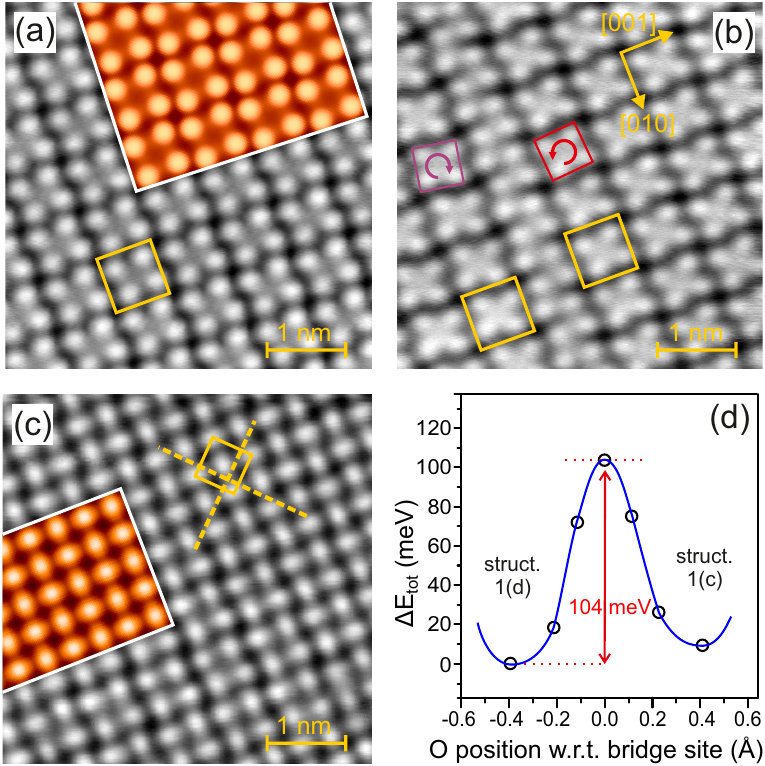}
	\caption{STM images (3.6\,nm$\times$3.6\,nm; $\pm20$\,mV; 1.5\,nA) of the 0.5\,ML oxygen phase on Rh(100) taken at 6\,K (a,b) and 78\,K (c) with the apparent translational symmetry indicated by yellow squares. 
	(a) Single domain of the $2\sqrt{2}$ phase together with the STM simulation (colored online, framed by a white line) of the model depicted in Fig.\,\ref{LEED_Models}(d). 
	(b) A domain boundary between two mirror domains, where oxygen atoms are arranged according to the model Fig.\,\ref{LEED_Models}(c). 
	(c) Apparent \twobytwo-periodicity with two glide symmetry planes (dashed yellow lines) with some $2\sqrt{2}$-like residues in the lower part. 
	Overlayed (colored online, framed by a white line) is an STM simulation for quasi-simultaneous site occupation.  
	(d) DFT calculation of the energy barrier between the two relaxed structures according to figures \ref{LEED_Models}(c) and \ref{LEED_Models}(d) that differ in the position of one oxygen atom. The blue solid line is a guide to the eye.}\label{STM_Results}
\end{figure}

Although this approach at first glance contradicts the symmetry and size of the unit cell as measured by LEED [Fig.\,\ref{LEED_Models}(a)], it is supported by very-low temperature STM images of the system ($T=6$\,K). Fig.\,\ref{STM_Results}(a) and \ref{STM_Results}(b) clearly show the presence of a $2\sqrt{2}$ unit cell (yellow squares). We can identify this phase as corresponding to the model of Fig.\,\ref{LEED_Models}(d) via the excellent agreement with a respective DFT-predicted STM image (employing the Tersoff-Hamann approximation \cite{Tersoff85}), which is overlaid on Fig.\,\ref{STM_Results}(a). 
This also proves that the bright protrusions imaged in STM are the oxygen atoms within the structure. 
Due to their off-center position at threefold-coordinated sites of the substrate's reconstruction, the quartets of oxygen atoms are slightly rotated against the crystallographic main axes, either clockwise or anti-clockwise, so that two mirror domains are possible. 
Fig.\,\ref{STM_Results}(b) displays the boundary between such two domains, which consists of a well-defined stripe, where every four oxygen atoms are arranged in a trapezoid-like configuration. 
This arrangement is the motif in the model of Fig.\,\ref{LEED_Models}(c). 
In contrast, a simple diagonal zig-zag arrangement of oxygen atoms, which would be characteristic of the conventional \twobytwo\ model, cannot be found anywhere on the surface. 

These experimental findings can be rationalized by comparing the total energies of the structural models of Figs.\,\ref{LEED_Models}(b-d) resulting from our DFT calculations, cf.\ Table \ref{DFTresults}: 
Independent of the functional chosen for the calculation, model (d) always turns out to be the configuration with the lowest total energy (set to zero), closely followed by model (c), which is only 3--9\,meV less favorable per $2\sqrt{2}$ cell. In contrast, the \twobytwo\ model (b) is about 20\,meV higher in energy. 
This means that the observed $2\sqrt{2}$ configuration is indeed the ground state structure of the 0.5\,ML O/Rh(100) system.

\begin{table}
	\caption{\label{DFTresults} Calculated total-energy difference $\Delta E$ in meV per $2\sqrt{2}$ cell of models Fig.\,\ref{LEED_Models}(b)+\ref{LEED_Models}(c)  with respect to the lowest energy configuration of model Fig.\,\ref{LEED_Models}(d) for various functionals.}
	\begin{ruledtabular}
		\begin{tabular}{cccc}
			model & PBE-PAW  & PBEsol  & optB86b\\
			& $\Delta E$ (meV)& $\Delta E$ (meV)&$\Delta E$ (meV) \\
			\hline
			Fig.\,\ref{LEED_Models}(b) & +16 & +28 & +23 \\	
			Fig.\,\ref{LEED_Models}(c) & +3 & +9 & +8 \\
			Fig.\,\ref{LEED_Models}(d) &  0 &  0 &  0 \\
			
		\end{tabular}
	\end{ruledtabular}
\end{table}

With respect to the local structure, all three models turn out to be rather similar in the DFT modeling. Rh atoms bound to two oxygen atoms always behave as if being larger in size compared to single-coordinated ones: They buckle out of the surface by about 0.1\,{\AA} and whenever possible they assume larger distances from each other. 
Thus, we have always two classes of structurally distinct Rh atoms within the topmost layer, and the three models only differ in the way how these atoms are mutually arranged. In Fig.\,\ref{LEED_Models}(b) they form linear chains, whereas in (c) they meander (zig-zag) and in (d) they form squares. 
Hence, it appears reasonable to correlate the energy differences with their ability to relax, considering the different effective atomic sizes of Rh atoms. 
This is certainly unfavorable in the case of the \twobytwo\ structure with its almost linear arrangement of atoms of the same type and best for the squares that can alternatingly expand and contract in a checkerboard pattern. 

 In STM the appearance of this phase has significantly changed at temperatures as low as 78\,K cf.\ Fig.\,\ref{STM_Results}(c). Over wide areas the oxygen atoms appear no more arranged as quartets but equally distant, however, with an elongated shape alternately lined up along [010] and [001] directions. So, in STM we ``see'' a \twobytwo\ mesh with p4gm symmetry. 
Since the alternating elongation of STM protrusions just coincides with the directions in which neighbored threefold adsorption sites are aligned, it seems obvious to interpret this finding as a quasi-simultaneous occupation of adjacent sites. For that, the oxygen atoms must change position in between these sites with a rate much faster than the scan speed of the STM tip.

This picture is supported by a DFT-based estimate of the diffusion barrier that determines the hopping rate of one single O atom within the $2\sqrt{2}$ cell, which switches the structure from model Fig.\,\ref{LEED_Models}(d) to Fig.\,\ref{LEED_Models}(c). To that end, we employed the climbing image nudge elastic band method (CI-NEB) \cite{Henkelman2000}.
The resulting minimum-energy path is shown in Fig.\,\ref{STM_Results}(d), where the barrier is about 100\,meV (PBEsol). Using an attempt frequency derived from a harmonic approximation to the energetic minima yields a rate of thermally activated hops of the order of megahertz at 78\,K. Thus, they are indeed much more frequent than the image acquisition (0.3\,ms/pixel). 
As a consequence, this STM image is a \textit{time average} of the surface configurations. 
The elongated atoms in Fig.\,\ref{STM_Results}(c) correspond exactly to the superposition of images of an oxygen atom taken at 6\,K on either side of the short-bridge barrier, which is visualized in a corresponding STM simulation overlayed in Fig.\,\ref{STM_Results}(c). 

From the STM image we cannot infer whether the oxygen atoms completely lose or locally retain their lateral order, i.e.\ we cannot decide whether they hop independent of each other or coherently, i.e their hops are associated with domain boundaries, letting them fluctuate across the surface. 
This, however, is possible by a quantitative LEED intensity analysis. 
Here, the scattering of electrons is instantaneous on the time scale of the hopping motion of the oxygen atoms so that the electrons just ``see'' snapshots of the adsorbate distribution. 
Moreover, the beam intensities are generated via multiple scattering processes on a spatial range of the electron's attenuation length, which is of the order of 10\,\AA\ and thus are very sensitive to the local order. 
We therefore fitted two alternative models to our experimental LEED-I(V) data taken at 100\,K: An ordered $2\sqrt{2}$ according to Fig.\,\ref{LEED_Models}(d) and a \twobytwo\ with both quasi-threefold oxygen adsorption sites being half-filled and the scattering between these two is neglected (to mimic statistical site occupation). 
For the latter model we achieve a best-fit R-factor of $R = 0.096$, quite comparable to the value of $R = 0.100$ obtained by us for the conventional \twobytwo\ model. 
In contrast, the $2\sqrt{2}$ model resulted in an ultra-low R-factor of $R = 0.073$ and, in combination with the large data base used for the analysis, an also very low variance of $var(R) = 0.005$. 
That means, we can rule out the other structural models with high statistical significance. 
A detailed discussion of the resulting best-fit structure and error margins for the fitted parameters is given elsewhere \cite{Hammer2025}, a structure file (POSCAR format) of the LEED best-fit is supplied as Supplementary Material \cite{SupMat}.

\begin{figure}[tb]
	\centering
	\includegraphics[width=0.9\columnwidth]{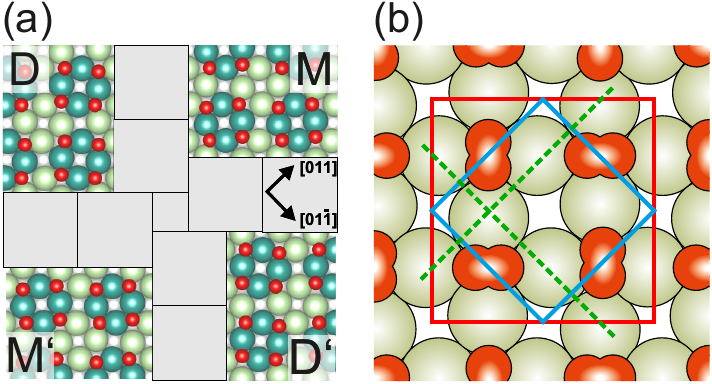}
	\caption{(a) At 100\,K, LEED will detect a momentary domain configuration as schematically shown. The gray $2\sqrt{2}$ areas are filled by the domain boundary structure Fig.\,\ref{LEED_Models}(c). The coherent superposition of the domains in (a) will produce a LEED pattern corresponding to the effective surface structure displayed in (b). This structure has a \twobytwo\ unit cell (blue square) with two orthogonal glide planes (green dashed lines).} 
	\label{coherence}
\end{figure}

At this point, where the presence of $2\sqrt{2}$ domains at the surface is evidenced also at temperatures around 100\,K, the reader might wonder why there are no characteristic quarter-order spots visible in the LEED pattern. 
Indeed, also our model calculations predict a sufficiently high intensity level for these spots to be easily detected. Their extinction in experiment can be explained by \textit{coherent} scattering effects as follows: 
We know already that the domain boundaries between mirror domains consist of single cells of the structure displayed in Fig.\,\ref{LEED_Models}(c) (cf.\ Fig.\,\ref{STM_Results}(b)), which cost only a few meV to create (Tab.\,\ref{DFTresults}).
Hence, at 78\,K they must be excited in large numbers.The structure decays into a multitude of nano-domains within the coherence area of the LEED experiment (typically 200\,\AA\ in diameter). 
We also see from Fig.\,\ref{STM_Results}(b) that mirror domains are shifted by half a $2\sqrt{2}$ unit vector against each other. 
Due to the fourfold symmetry of the structure we obtain domain boundaries in two perpendicular directions producing a total of four differently aligned domains D, D', M, and M' shown in Fig.\,\ref{coherence}(a). 
Thereby D' and M' are identical to D and M, but laterally shifted by half a cell diagonal, i.e.\ they are perfect anti-phase domains. 
Since the underlying \twobytwo\ `clock'-reconstruction of the substrate is mono-domain over very large distances, all $2\sqrt{2}$ domains of the same type are exactly in phase and thus would produce sharp spots on their own like a continuous grid with a fractional site occupancy. 
Additionally, we have to regard the \textit{interference} of scattered waves from all four domains. 
In particular, the quarter-order beams produced by the domains D and D' (or M and M') are scattered strictly anti-phase and because of equal statistical weights this leads to a complete extinction via destructive interference. The whole scenario can be modeled by a projection of all fractional scatterers (each with the respective lateral shift) on to the same unit cell as shown in Fig.\,\ref{coherence}(b). It clearly exhibits a higher symmetry than each of the single domains). This configuration can be described by a smaller \twobytwo\ unit mesh and it also possesses two perpendicular glide planes indicated by dashed lines (a mathematical proof is given elsewhere \cite{Hammer2025}) -- exactly what is observed in the LEED pattern. 
In fact, with extreme contrast enhancement we find very weak and diffuse quarter-order spots originating from short-range order effects indicating an average size of about 18\,\AA\ for the nano-domains in perfect agreement with corresponding kinematic diffraction simulations. (For details see ref.\,\cite{Hammer2025}.)

Finally, we want to emphasize that the knowledge about the correct surface structure is not an end in itself, but it is important for a microscopic understanding of surface-related processes like adsorption or catalytic reactions. 
Considering, e.g., co-adsorption of a CO molecule, then the formerly accepted \twobytwo\ model solely offers one type of fourfold-coordinated hollow sites, where two Rh atoms are singly and two are double-coordinated to oxygen, cf.\ Fig.\,\ref{LEED_Models}(b). DFT predicts that a CO molecule adsorbing on such a site would strongly distort the structure and eventually relax towards a bridge site. In contrast, the true $2\sqrt{2}$ structure additionally exhibits two further types of fourfold hollow sites, one with exclusively single and one with double O-coordinated Rh atoms (cf.\ Fig.\,\ref{LEED_Models}(d)). The former are the preferred adsorption sites. 
The calculated adsorption energies differ by more than 1\,eV among these sites, see Tab.\ref{adsorption_energies}. In particular, the all-single O-coordinated hollow site (square of light-colored atoms in Fig.\,\ref{LEED_Models}(d)) is found to be by 0.34\,eV more favorable than the mixed O-coordinated site offered by the \twobytwo\ phase. 
Certainly, these very different adsorption energies together with the varying local atomic environment will also drastically affect the activation barrier for CO oxidation at the surface. 
In practice, for the present system the CO oxidation reaction usually runs at much higher temperatures. 
One might argue that the order-dependent availability of different sites might be washed out by increasing disorder, but the large differences of CO adsorption energies (cf.\ Tab.\ \ref{adsorption_energies}) imply that CO will create at least locally a $2\sqrt{2}$ structure. In addition, one should note that any attempt to do a DFT calculation for the (wrong) $(2\times 2)$ unit cell will lead to substantial deviations of the adsorption energy. Therefore, this system serves as a prime example for how strongly local ordering effects can influence the chemical properties of a system. 
There is no microscopic understanding of the catalytic action of such a system without the unambiguous identification of available adsorption sites.

\begin{table}
	\caption{\label{adsorption_energies} Calculated adsorption energy differences  $\Delta E_{ad}$ per $2\sqrt{2}$ cell for a CO molecule placed on different fourfold hollow sites of models Fig.\,\ref{LEED_Models}(b)+\ref{LEED_Models}(d). The sites differ by the O-coordination of involved Rh atoms (solely 1- or 2-fold, or 1,2-mixed). }
	\renewcommand{\arraystretch}{1.4}
	\begin{ruledtabular}
		\begin{tabular}{cccc}
			structure & O-coord. of Rh & adsorption site  & energy [meV]  \\ 
			\hline

			\twobytwo\ & 1,2-mixed & moved to bridge & +344  \\
			$2\sqrt{2}$    & 1-fold    & 4-fold hollow   & 0     \\
					   & 1,2-mixed & moved to bridge & +186  \\			
			           & 2-fold    & 4-fold hollow & +1077  \\			
			
		\end{tabular}
	\end{ruledtabular}
\end{table}

In conclusion, we have identified the true ground-state structure of the 0.5\,ML oxygen phase on Rh(100), which is effectively hidden from recognition by standard structure-sensitive methods like qualitative LEED or STM. This is because already at 100\,K this phase  breaks down into an ensemble of rapidly fluctuating nanodomains. The superposition of their contributions to any method fakes both the apparent periodicity and local symmetry and hence, inevitably leads to wrong model structures. 
Consequently, also other physical or chemical properties of the surface like e.g.\ adsorption sites or reaction barriers derived from the wrong structure model will be inaccurate or invalid. 
The prerequisites for such a scenario are a low formation energy for domain boundaries and a high hopping rate of adatoms at the temperature of investigation. 
As these requirements are fulfilled in a large number of systems this could lead to similar problems that have not yet been discovered.

\vspace{5mm}

This research was funded in part by the Austrian Science Fund (FWF) [Grant 10.55776/F8100, SFB Project TACO] and the Deutsche Forschungsgemeinschaft (DFG) [Research Unit FOR 1878 ``\textit{fun}COS'']. For open access purposes, the authors have applied a CC BY public copyright license to any author accepted manuscript version arising from this submission. The authors acknowledge TU Wien Bibliothek for financial support through its Open Access Funding Programme.
The authors gratefully acknowledge the scientific support and HPC resources provided by the Erlangen National High Performance Computing Center (NHR@FAU) of the Friedrich-Alexander-Universit\"{a}t Erlangen-N\"{u}rnberg (FAU).

\appendix*
\section{Experimental Procedures and Methods}
Here we briefly describe the experimental and computational procedures applied in this study. 
All shown LEED data were acquired in Erlangen, where also the LEED-I(V) and DFT calculations were performed as well as the short-range order simulations. All low-temperature STM data were taken in Vienna. 
For more detailed information see ref.\,\cite{Hammer2025}.

\subsection{Experimental details}
All experiments were carried out in ultra-high vacuum (UHV). 
\\
The Rh(100) surface was prepared by several cleaning cycles each consisting of subsequent Ar$^+$ bombardment, annealing at 870\,K in $1 \cdot 10^{-6}$\,mbar O$_2$ and final flash to 1090\,K in UHV. The preparation of the $(2\sqrt{2}$ phase of oxygen is already described in the main text.
\\
The UHV chamber in Erlangen (base pressure $\approx 2 \cdot 10^{-11}$\,mbar) housing an ERLEED optics and a room-temperature beetle type STM is described in \cite{Kisslinger2017}.  
In this system the sample could be heated rapidly to temperatures up to $1400~\mbox{K}$ by electron bombardment from the rear and cooled to about $100~\mbox{K}$ by direct contact to a liquid nitrogen reservoir within minutes. 
The temperature was measured by a Type-K thermocouple spot-welded directly onto the rim of the crystal. The sample holder allowed for independent sample rotation around two orthogonal axes lying both within the surface plane so that normal incidence of the electron beam could be precisely aligned for LEED-I(V) data acquisition (accuracy $\approx 0.1^\circ$). To minimize both the influence of residual gas adsorption and the thermal diffuse scattering background, data acquisition was started immediately after the temperature reached $120~\mbox{K}$ in the cool-down process. The LEED pattern was recorded by a CCD camera between 40--600$~\mbox{eV}$ in $0.5~\mbox{eV}$ steps and stored on a computer for later off-line evaluation. 
The I(V) spectra were averaged between symmetrically equivalent beams, slightly smoothed for noise removal and normalized by the simultaneously recorded primary beam current. This resulted in a database for the LEED analysis of $\Delta E \approx 7000~\mbox{eV}$.
\\
The experiments in Vienna were also performed in a two-chamber UHV system consisting of a similarly equipped preparation chamber (base pressure  $7 \cdot 10^{-11}$\,mbar) and an adjacent analysis chamber (base pressure $5 \cdot 10^{-12}$\,mbar). The analysis chamber hosts a low-temperature STM (Omicron LT-STM) operating at 79\,K and 6\,K using electrochemically etched W tips. In this system, annealing temperatures were measured by a thermocouple on the non-transferable part of the sample holder, which leads to estimated errors of $\approx{50}$\,K at the annealing temperature.

All shown LEED data were acquired in Erlangen, where also the LEED-I(V) and DFT calculations were performed as well as the short-range order simulations. All low-temperature STM data were taken in Vienna.

\subsection{Computational details}
Full-dynamically calculated model intensities as well the optimization of model parameters by means of \textit{tensor LEED} were performed using the \textsc{ViPERLEED} package \cite{ViPErLEED_I} implementing the phase shift program \textsc{EASiSSS} \cite{Rundgren2003} and the \textsc{TensErLEED} program package \cite{Blum2001}. 
For quantification of the agreement between experimental and theoretical spectra the Pendry R-factor $R$ \cite{Pendry1980} was used.
\\
In the LEED analysis we varied all geometrical parameters of oxygen and rhodium atoms of the first two layers as far as allowed by symmetry, as well as average layer distances of the third and forth Rh layers.
Additionally, vibrational amplitudes were varied for the oxygen atoms and the top-layer Rh atoms as well as a constant shift of the experimental energies to account for the work function of the cathode and the unknown reference value of the calculated inner potential curve \cite{Rundgren2003}. 
This resulted in a total of $P = 23$ free parameters (17 structural and 6 non-structural), which have been adjusted.  The fit can be regarded as highly reliable because of the large I(V)-data base used
(redundancy factor $\rho =  \Delta E / 4V_{0i} P \approx 15.7$).

All DFT calculations were performed using the VASP package \cite{vasp3} employing the PBE \cite{PBE} general gradient approximation and its refinement PBEsol \cite{PBEsol}. For comparison, calculations were also performed with the optB86b vdW-functional of Klime\v{s} et al.\,\cite{optB86b-vdW}. 
Nine-layer Rh(100)-$2\sqrt{2}$ slabs with the center layer fixed to the bulk coordinates and a 15\,\AA\ vacuum gap were set up. 
Oxygen atoms were positioned on one surface of the slab only.
The calculations were performed with an energy cutoff of 550\,eV employing an automatically generated $10 \times 10 \times 1$ Monkhorst k-point mesh (50 irreducible k-points). 
Convergence tests revealed that at those settings energies are reliable to below 5\,meV.
For the climbing image nudged elastic band (CI-NEB) method we used the implementation of the Henkelman group \cite{Henkelman2000,VTSTtools}.
For structural comparison all calculated results were scaled by the ratio of the bulk Rh lattice constant from experiment (3.80\,\AA) and that obtained from bulk calculations of Rh employing the respective functionals.
Lattice parameters PBE: 3.83\,\AA\ (same as ref.\,\cite{Scheffler1999}), PBEsol: 3.78\,\AA, optB86b: 3.80\,\AA.

\bibliography{literatur_Rh-O}

\begin{thebibliography}{33}%
\makeatletter
\providecommand \@ifxundefined [1]{%
 \@ifx{#1\undefined}
}%
\providecommand \@ifnum [1]{%
 \ifnum #1\expandafter \@firstoftwo
 \else \expandafter \@secondoftwo
 \fi
}%
\providecommand \@ifx [1]{%
 \ifx #1\expandafter \@firstoftwo
 \else \expandafter \@secondoftwo
 \fi
}%
\providecommand \natexlab [1]{#1}%
\providecommand \enquote  [1]{``#1''}%
\providecommand \bibnamefont  [1]{#1}%
\providecommand \bibfnamefont [1]{#1}%
\providecommand \citenamefont [1]{#1}%
\providecommand \href@noop [0]{\@secondoftwo}%
\providecommand \href [0]{\begingroup \@sanitize@url \@href}%
\providecommand \@href[1]{\@@startlink{#1}\@@href}%
\providecommand \@@href[1]{\endgroup#1\@@endlink}%
\providecommand \@sanitize@url [0]{\catcode `\\12\catcode `\$12\catcode
  `\&12\catcode `\#12\catcode `\^12\catcode `\_12\catcode `\%12\relax}%
\providecommand \@@startlink[1]{}%
\providecommand \@@endlink[0]{}%
\providecommand \url  [0]{\begingroup\@sanitize@url \@url }%
\providecommand \@url [1]{\endgroup\@href {#1}{\urlprefix }}%
\providecommand \urlprefix  [0]{URL }%
\providecommand \Eprint [0]{\href }%
\providecommand \doibase [0]{https://doi.org/}%
\providecommand \selectlanguage [0]{\@gobble}%
\providecommand \bibinfo  [0]{\@secondoftwo}%
\providecommand \bibfield  [0]{\@secondoftwo}%
\providecommand \translation [1]{[#1]}%
\providecommand \BibitemOpen [0]{}%
\providecommand \bibitemStop [0]{}%
\providecommand \bibitemNoStop [0]{.\EOS\space}%
\providecommand \EOS [0]{\spacefactor3000\relax}%
\providecommand \BibitemShut  [1]{\csname bibitem#1\endcsname}%
\let\auto@bib@innerbib\@empty
%</preamble>
\bibitem [{\citenamefont {Somorjai}\ and\ \citenamefont
  {Van~Hove}(1995)}]{Somorjai1995}%
  \BibitemOpen
  \bibfield  {author} {\bibinfo {author} {\bibfnamefont {G.~A.}\ \bibnamefont
  {Somorjai}}\ and\ \bibinfo {author} {\bibfnamefont {M.~A.}\ \bibnamefont
  {Van~Hove}},\ }\href {https://doi.org/10.1107/S0108768195001078} {\bibfield
  {journal} {\bibinfo  {journal} {Acta Crystallographica B}\ }\textbf {\bibinfo
  {volume} {51}},\ \bibinfo {pages} {502} (\bibinfo {year} {1995})}\BibitemShut
  {NoStop}%
\bibitem [{\citenamefont {Tan}\ \emph {et~al.}(2018)\citenamefont {Tan},
  \citenamefont {Huang}, \citenamefont {Liu},\ and\ \citenamefont
  {Wang}}]{Tan2018}%
  \BibitemOpen
  \bibfield  {author} {\bibinfo {author} {\bibfnamefont {L.}~\bibnamefont
  {Tan}}, \bibinfo {author} {\bibfnamefont {L.}~\bibnamefont {Huang}}, \bibinfo
  {author} {\bibfnamefont {Y.}~\bibnamefont {Liu}},\ and\ \bibinfo {author}
  {\bibfnamefont {Q.}~\bibnamefont {Wang}},\ }\href
  {https://doi.org/10.1021/acs.langmuir.7b04383} {\bibfield  {journal}
  {\bibinfo  {journal} {Langmuir}\ }\textbf {\bibinfo {volume} {34}},\ \bibinfo
  {pages} {5174} (\bibinfo {year} {2018})}\BibitemShut {NoStop}%
\bibitem [{\citenamefont {Shelef}\ and\ \citenamefont
  {Graham}(1994)}]{Shelef1994}%
  \BibitemOpen
  \bibfield  {author} {\bibinfo {author} {\bibfnamefont {M.}~\bibnamefont
  {Shelef}}\ and\ \bibinfo {author} {\bibfnamefont {G.~W.}\ \bibnamefont
  {Graham}},\ }\href {https://doi.org/10.1080/01614949408009468} {\bibfield
  {journal} {\bibinfo  {journal} {Catalysis Reviews}\ }\textbf {\bibinfo
  {volume} {36}},\ \bibinfo {pages} {433} (\bibinfo {year} {1994})}\BibitemShut
  {NoStop}%
\bibitem [{\citenamefont {Davda}\ \emph {et~al.}(2005)\citenamefont {Davda},
  \citenamefont {Shabaker}, \citenamefont {Huber}, \citenamefont {Cortright},\
  and\ \citenamefont {Dumesic}}]{Davda2005}%
  \BibitemOpen
  \bibfield  {author} {\bibinfo {author} {\bibfnamefont {R.}~\bibnamefont
  {Davda}}, \bibinfo {author} {\bibfnamefont {J.}~\bibnamefont {Shabaker}},
  \bibinfo {author} {\bibfnamefont {G.}~\bibnamefont {Huber}}, \bibinfo
  {author} {\bibfnamefont {R.}~\bibnamefont {Cortright}},\ and\ \bibinfo
  {author} {\bibfnamefont {J.}~\bibnamefont {Dumesic}},\ }\href
  {https://doi.org/https://doi.org/10.1016/j.apcatb.2004.04.027} {\bibfield
  {journal} {\bibinfo  {journal} {Applied Catalysis B: Environmental}\ }\textbf
  {\bibinfo {volume} {56}},\ \bibinfo {pages} {171 } (\bibinfo {year}
  {2005})}\BibitemShut {NoStop}%
\bibitem [{\citenamefont {Gustafson}\ \emph {et~al.}(2014)\citenamefont
  {Gustafson}, \citenamefont {Blomberg}, \citenamefont {Martin}, \citenamefont
  {Fernandes}, \citenamefont {Borg}, \citenamefont {Liu}, \citenamefont
  {Chang},\ and\ \citenamefont {Lundgren}}]{Gustafson2013}%
  \BibitemOpen
  \bibfield  {author} {\bibinfo {author} {\bibfnamefont {J.}~\bibnamefont
  {Gustafson}}, \bibinfo {author} {\bibfnamefont {S.}~\bibnamefont {Blomberg}},
  \bibinfo {author} {\bibfnamefont {N.~M.}\ \bibnamefont {Martin}}, \bibinfo
  {author} {\bibfnamefont {V.}~\bibnamefont {Fernandes}}, \bibinfo {author}
  {\bibfnamefont {A.}~\bibnamefont {Borg}}, \bibinfo {author} {\bibfnamefont
  {Z.}~\bibnamefont {Liu}}, \bibinfo {author} {\bibfnamefont {R.}~\bibnamefont
  {Chang}},\ and\ \bibinfo {author} {\bibfnamefont {E.}~\bibnamefont
  {Lundgren}},\ }\href {https://doi.org/10.1088/0953-8984/26/5/055003}
  {\bibfield  {journal} {\bibinfo  {journal} {J. Phys. Condens. Matter}\
  }\textbf {\bibinfo {volume} {26}},\ \bibinfo {pages} {055003} (\bibinfo
  {year} {2014})}\BibitemShut {NoStop}%
\bibitem [{\citenamefont {Gustafson}\ \emph {et~al.}(2012)\citenamefont
  {Gustafson}, \citenamefont {Lundgren}, \citenamefont {Mikkelsen},
  \citenamefont {Borg}, \citenamefont {Klikovits}, \citenamefont {Schmid},
  \citenamefont {Varga},\ and\ \citenamefont {Andersen}}]{Gustafson2012}%
  \BibitemOpen
  \bibfield  {author} {\bibinfo {author} {\bibfnamefont {J.}~\bibnamefont
  {Gustafson}}, \bibinfo {author} {\bibfnamefont {E.}~\bibnamefont {Lundgren}},
  \bibinfo {author} {\bibfnamefont {A.}~\bibnamefont {Mikkelsen}}, \bibinfo
  {author} {\bibfnamefont {M.}~\bibnamefont {Borg}}, \bibinfo {author}
  {\bibfnamefont {J.}~\bibnamefont {Klikovits}}, \bibinfo {author}
  {\bibfnamefont {M.}~\bibnamefont {Schmid}}, \bibinfo {author} {\bibfnamefont
  {P.}~\bibnamefont {Varga}},\ and\ \bibinfo {author} {\bibfnamefont {J.~N.}\
  \bibnamefont {Andersen}},\ }\href
  {http://stacks.iop.org/0953-8984/24/i=22/a=225006} {\bibfield  {journal}
  {\bibinfo  {journal} {J. Phys. Condens. Matter}\ }\textbf {\bibinfo {volume}
  {24}},\ \bibinfo {pages} {225006} (\bibinfo {year} {2012})}\BibitemShut
  {NoStop}%
\bibitem [{\citenamefont {Ki{\ss}linger}\ \emph {et~al.}(2017)\citenamefont
  {Ki{\ss}linger}, \citenamefont {Ferstl}, \citenamefont {Schneider},\ and\
  \citenamefont {Hammer}}]{Kisslinger2017}%
  \BibitemOpen
  \bibfield  {author} {\bibinfo {author} {\bibfnamefont {T.}~\bibnamefont
  {Ki{\ss}linger}}, \bibinfo {author} {\bibfnamefont {P.}~\bibnamefont
  {Ferstl}}, \bibinfo {author} {\bibfnamefont {M.~A.}\ \bibnamefont
  {Schneider}},\ and\ \bibinfo {author} {\bibfnamefont {L.}~\bibnamefont
  {Hammer}},\ }\href {https://doi.org/10.1088/1361-648x/aa7db7} {\bibfield
  {journal} {\bibinfo  {journal} {J. Phys.: Condens. Matter}\ }\textbf
  {\bibinfo {volume} {29}},\ \bibinfo {pages} {365001} (\bibinfo {year}
  {2017})}\BibitemShut {NoStop}%
\bibitem [{\citenamefont {Gustafson}\ \emph {et~al.}(2005)\citenamefont
  {Gustafson}, \citenamefont {Mikkelsen}, \citenamefont {Borg}, \citenamefont
  {Andersen}, \citenamefont {Lundgren}, \citenamefont {Klein}, \citenamefont
  {Hofer}, \citenamefont {Schmid}, \citenamefont {Varga}, \citenamefont
  {K\"ohler}, \citenamefont {Kresse}, \citenamefont {Kasper}, \citenamefont
  {Stierle},\ and\ \citenamefont {Dosch}}]{Gustafson2005}%
  \BibitemOpen
  \bibfield  {author} {\bibinfo {author} {\bibfnamefont {J.}~\bibnamefont
  {Gustafson}}, \bibinfo {author} {\bibfnamefont {A.}~\bibnamefont
  {Mikkelsen}}, \bibinfo {author} {\bibfnamefont {M.}~\bibnamefont {Borg}},
  \bibinfo {author} {\bibfnamefont {J.~N.}\ \bibnamefont {Andersen}}, \bibinfo
  {author} {\bibfnamefont {E.}~\bibnamefont {Lundgren}}, \bibinfo {author}
  {\bibfnamefont {C.}~\bibnamefont {Klein}}, \bibinfo {author} {\bibfnamefont
  {W.}~\bibnamefont {Hofer}}, \bibinfo {author} {\bibfnamefont
  {M.}~\bibnamefont {Schmid}}, \bibinfo {author} {\bibfnamefont
  {P.}~\bibnamefont {Varga}}, \bibinfo {author} {\bibfnamefont
  {L.}~\bibnamefont {K\"ohler}}, \bibinfo {author} {\bibfnamefont
  {G.}~\bibnamefont {Kresse}}, \bibinfo {author} {\bibfnamefont
  {N.}~\bibnamefont {Kasper}}, \bibinfo {author} {\bibfnamefont
  {A.}~\bibnamefont {Stierle}},\ and\ \bibinfo {author} {\bibfnamefont
  {H.}~\bibnamefont {Dosch}},\ }\href
  {https://doi.org/10.1103/PhysRevB.71.115442} {\bibfield  {journal} {\bibinfo
  {journal} {Phys. Rev. B}\ }\textbf {\bibinfo {volume} {71}},\ \bibinfo
  {pages} {115442} (\bibinfo {year} {2005})}\BibitemShut {NoStop}%
\bibitem [{\citenamefont {Oed}\ \emph {et~al.}(1988)\citenamefont {Oed},
  \citenamefont {D{\"o}tsch}, \citenamefont {Hammer}, \citenamefont {Heinz},\
  and\ \citenamefont {M{\"u}ller}}]{Oed1988}%
  \BibitemOpen
  \bibfield  {author} {\bibinfo {author} {\bibfnamefont {W.}~\bibnamefont
  {Oed}}, \bibinfo {author} {\bibfnamefont {B.}~\bibnamefont {D{\"o}tsch}},
  \bibinfo {author} {\bibfnamefont {L.}~\bibnamefont {Hammer}}, \bibinfo
  {author} {\bibfnamefont {K.}~\bibnamefont {Heinz}},\ and\ \bibinfo {author}
  {\bibfnamefont {K.}~\bibnamefont {M{\"u}ller}},\ }\href
  {https://doi.org/http://dx.doi.org/10.1016/0039-6028(88)90246-4} {\bibfield
  {journal} {\bibinfo  {journal} {Surf. Sci.}\ }\textbf {\bibinfo {volume}
  {207}},\ \bibinfo {pages} {55} (\bibinfo {year} {1988})}\BibitemShut
  {NoStop}%
\bibitem [{\citenamefont {Mercer}\ \emph {et~al.}(1996)\citenamefont {Mercer},
  \citenamefont {Finetti}, \citenamefont {Leibsle}, \citenamefont {McGrath},
  \citenamefont {Dhanak}, \citenamefont {Baraldi}, \citenamefont {Prince},\
  and\ \citenamefont {Rosei}}]{Mercer1996}%
  \BibitemOpen
  \bibfield  {author} {\bibinfo {author} {\bibfnamefont {J.}~\bibnamefont
  {Mercer}}, \bibinfo {author} {\bibfnamefont {P.}~\bibnamefont {Finetti}},
  \bibinfo {author} {\bibfnamefont {F.}~\bibnamefont {Leibsle}}, \bibinfo
  {author} {\bibfnamefont {R.}~\bibnamefont {McGrath}}, \bibinfo {author}
  {\bibfnamefont {V.}~\bibnamefont {Dhanak}}, \bibinfo {author} {\bibfnamefont
  {A.}~\bibnamefont {Baraldi}}, \bibinfo {author} {\bibfnamefont
  {K.}~\bibnamefont {Prince}},\ and\ \bibinfo {author} {\bibfnamefont
  {R.}~\bibnamefont {Rosei}},\ }\href
  {https://doi.org/https://doi.org/10.1016/0039-6028(95)01126-9} {\bibfield
  {journal} {\bibinfo  {journal} {Surf. Sci.}\ }\textbf {\bibinfo {volume}
  {352-354}},\ \bibinfo {pages} {173 } (\bibinfo {year} {1996})}\BibitemShut
  {NoStop}%
\bibitem [{\citenamefont {Baraldi}\ \emph {et~al.}(1997)\citenamefont
  {Baraldi}, \citenamefont {Dhanak}, \citenamefont {Comelli}, \citenamefont
  {Prince},\ and\ \citenamefont {Rosei}}]{Baraldi1997}%
  \BibitemOpen
  \bibfield  {author} {\bibinfo {author} {\bibfnamefont {A.}~\bibnamefont
  {Baraldi}}, \bibinfo {author} {\bibfnamefont {V.~R.}\ \bibnamefont {Dhanak}},
  \bibinfo {author} {\bibfnamefont {G.}~\bibnamefont {Comelli}}, \bibinfo
  {author} {\bibfnamefont {K.~C.}\ \bibnamefont {Prince}},\ and\ \bibinfo
  {author} {\bibfnamefont {R.}~\bibnamefont {Rosei}},\ }\href
  {https://doi.org/10.1103/PhysRevB.56.10511} {\bibfield  {journal} {\bibinfo
  {journal} {Phys. Rev. B}\ }\textbf {\bibinfo {volume} {56}},\ \bibinfo
  {pages} {10511} (\bibinfo {year} {1997})}\BibitemShut {NoStop}%
\bibitem [{\citenamefont {Shen}\ \emph {et~al.}(1998)\citenamefont {Shen},
  \citenamefont {Qayyum}, \citenamefont {O'Connor},\ and\ \citenamefont
  {King}}]{Shen1998}%
  \BibitemOpen
  \bibfield  {author} {\bibinfo {author} {\bibfnamefont {Y.~G.}\ \bibnamefont
  {Shen}}, \bibinfo {author} {\bibfnamefont {A.}~\bibnamefont {Qayyum}},
  \bibinfo {author} {\bibfnamefont {D.~J.}\ \bibnamefont {O'Connor}},\ and\
  \bibinfo {author} {\bibfnamefont {B.~V.}\ \bibnamefont {King}},\ }\href
  {https://doi.org/10.1103/PhysRevB.58.10025} {\bibfield  {journal} {\bibinfo
  {journal} {Phys. Rev. B}\ }\textbf {\bibinfo {volume} {58}},\ \bibinfo
  {pages} {10025} (\bibinfo {year} {1998})}\BibitemShut {NoStop}%
\bibitem [{\citenamefont {Alfe}\ \emph {et~al.}(1998)\citenamefont {Alfe},
  \citenamefont {de~Gironcoli},\ and\ \citenamefont {Baroni}}]{Alfe1998}%
  \BibitemOpen
  \bibfield  {author} {\bibinfo {author} {\bibfnamefont {D.}~\bibnamefont
  {Alfe}}, \bibinfo {author} {\bibfnamefont {S.}~\bibnamefont {de~Gironcoli}},\
  and\ \bibinfo {author} {\bibfnamefont {S.}~\bibnamefont {Baroni}},\ }\href
  {https://doi.org/https://doi.org/10.1016/S0039-6028(98)00120-4} {\bibfield
  {journal} {\bibinfo  {journal} {Surface Science}\ }\textbf {\bibinfo {volume}
  {410}},\ \bibinfo {pages} {151 } (\bibinfo {year} {1998})}\BibitemShut
  {NoStop}%
\bibitem [{\citenamefont {Baraldi}\ \emph {et~al.}(1999)\citenamefont
  {Baraldi}, \citenamefont {Cerd\'a}, \citenamefont {Mart\'{\i}n-Gago},
  \citenamefont {Comelli}, \citenamefont {Lizzit}, \citenamefont {Paolucci},\
  and\ \citenamefont {Rosei}}]{Baraldi1999}%
  \BibitemOpen
  \bibfield  {author} {\bibinfo {author} {\bibfnamefont {A.}~\bibnamefont
  {Baraldi}}, \bibinfo {author} {\bibfnamefont {J.}~\bibnamefont {Cerd\'a}},
  \bibinfo {author} {\bibfnamefont {J.~A.}\ \bibnamefont {Mart\'{\i}n-Gago}},
  \bibinfo {author} {\bibfnamefont {G.}~\bibnamefont {Comelli}}, \bibinfo
  {author} {\bibfnamefont {S.}~\bibnamefont {Lizzit}}, \bibinfo {author}
  {\bibfnamefont {G.}~\bibnamefont {Paolucci}},\ and\ \bibinfo {author}
  {\bibfnamefont {R.}~\bibnamefont {Rosei}},\ }\href
  {https://doi.org/10.1103/PhysRevLett.82.4874} {\bibfield  {journal} {\bibinfo
   {journal} {Phys. Rev. Lett.}\ }\textbf {\bibinfo {volume} {82}},\ \bibinfo
  {pages} {4874} (\bibinfo {year} {1999})}\BibitemShut {NoStop}%
\bibitem [{\citenamefont {Alfè}\ \emph {et~al.}(1999)\citenamefont {Alfè},
  \citenamefont {de~Gironcoli},\ and\ \citenamefont {Baroni}}]{Alfe1999}%
  \BibitemOpen
  \bibfield  {author} {\bibinfo {author} {\bibfnamefont {D.}~\bibnamefont
  {Alfè}}, \bibinfo {author} {\bibfnamefont {S.}~\bibnamefont
  {de~Gironcoli}},\ and\ \bibinfo {author} {\bibfnamefont {S.}~\bibnamefont
  {Baroni}},\ }\href
  {https://doi.org/https://doi.org/10.1016/S0039-6028(99)00620-2} {\bibfield
  {journal} {\bibinfo  {journal} {Surf. Sci.}\ }\textbf {\bibinfo {volume}
  {437}},\ \bibinfo {pages} {18 } (\bibinfo {year} {1999})}\BibitemShut
  {NoStop}%
\bibitem [{\citenamefont {Norris}\ \emph {et~al.}(2000)\citenamefont {Norris},
  \citenamefont {Schedin}, \citenamefont {Thornton}, \citenamefont {Dhanak},
  \citenamefont {Turner},\ and\ \citenamefont {McGrath}}]{Norris2000}%
  \BibitemOpen
  \bibfield  {author} {\bibinfo {author} {\bibfnamefont {A.~G.}\ \bibnamefont
  {Norris}}, \bibinfo {author} {\bibfnamefont {F.}~\bibnamefont {Schedin}},
  \bibinfo {author} {\bibfnamefont {G.}~\bibnamefont {Thornton}}, \bibinfo
  {author} {\bibfnamefont {V.~R.}\ \bibnamefont {Dhanak}}, \bibinfo {author}
  {\bibfnamefont {T.~S.}\ \bibnamefont {Turner}},\ and\ \bibinfo {author}
  {\bibfnamefont {R.}~\bibnamefont {McGrath}},\ }\href
  {https://doi.org/10.1103/PhysRevB.62.2113} {\bibfield  {journal} {\bibinfo
  {journal} {Phys. Rev. B}\ }\textbf {\bibinfo {volume} {62}},\ \bibinfo
  {pages} {2113} (\bibinfo {year} {2000})}\BibitemShut {NoStop}%
\bibitem [{\citenamefont {Kirsch}\ and\ \citenamefont
  {Harris}(2004)}]{Kirsch2004}%
  \BibitemOpen
  \bibfield  {author} {\bibinfo {author} {\bibfnamefont {J.~E.}\ \bibnamefont
  {Kirsch}}\ and\ \bibinfo {author} {\bibfnamefont {S.}~\bibnamefont
  {Harris}},\ }\href
  {https://doi.org/https://doi.org/10.1016/j.susc.2004.01.037} {\bibfield
  {journal} {\bibinfo  {journal} {Surf. Sci.}\ }\textbf {\bibinfo {volume}
  {553}},\ \bibinfo {pages} {82} (\bibinfo {year} {2004})}\BibitemShut
  {NoStop}%
\bibitem [{\citenamefont {Bianchettin}\ \emph {et~al.}(2009)\citenamefont
  {Bianchettin}, \citenamefont {Baraldi}, \citenamefont {de~Gironcoli},
  \citenamefont {Vesselli}, \citenamefont {Lizzit}, \citenamefont {Comelli},\
  and\ \citenamefont {Rosei}}]{Bianchettin2009}%
  \BibitemOpen
  \bibfield  {author} {\bibinfo {author} {\bibfnamefont {L.}~\bibnamefont
  {Bianchettin}}, \bibinfo {author} {\bibfnamefont {A.}~\bibnamefont
  {Baraldi}}, \bibinfo {author} {\bibfnamefont {S.}~\bibnamefont
  {de~Gironcoli}}, \bibinfo {author} {\bibfnamefont {E.}~\bibnamefont
  {Vesselli}}, \bibinfo {author} {\bibfnamefont {S.}~\bibnamefont {Lizzit}},
  \bibinfo {author} {\bibfnamefont {G.}~\bibnamefont {Comelli}},\ and\ \bibinfo
  {author} {\bibfnamefont {R.}~\bibnamefont {Rosei}},\ }\href
  {https://doi.org/10.1021/jp901223d} {\bibfield  {journal} {\bibinfo
  {journal} {J. Phys. Chem. C}\ }\textbf {\bibinfo {volume} {113}},\ \bibinfo
  {pages} {13192} (\bibinfo {year} {2009})}\BibitemShut {NoStop}%
\bibitem [{\citenamefont {Tan}\ \emph {et~al.}(2017)\citenamefont {Tan},
  \citenamefont {Huang}, \citenamefont {Wang},\ and\ \citenamefont
  {Liu}}]{Tan2017}%
  \BibitemOpen
  \bibfield  {author} {\bibinfo {author} {\bibfnamefont {L.}~\bibnamefont
  {Tan}}, \bibinfo {author} {\bibfnamefont {L.}~\bibnamefont {Huang}}, \bibinfo
  {author} {\bibfnamefont {Q.}~\bibnamefont {Wang}},\ and\ \bibinfo {author}
  {\bibfnamefont {Y.}~\bibnamefont {Liu}},\ }\href
  {https://doi.org/10.1021/acs.langmuir.7b02030} {\bibfield  {journal}
  {\bibinfo  {journal} {Langmuir}\ }\textbf {\bibinfo {volume} {33}},\ \bibinfo
  {pages} {11156} (\bibinfo {year} {2017})}\BibitemShut {NoStop}%
\bibitem [{\citenamefont {Hammer}\ \emph {et~al.}()\citenamefont {Hammer},
  \citenamefont {Kißlinger}, \citenamefont {Wagner}, \citenamefont {Neder},
  \citenamefont {Schmid}, \citenamefont {Diebold},\ and\ \citenamefont
  {Schneider}}]{Hammer2025}%
  \BibitemOpen
  \bibfield  {author} {\bibinfo {author} {\bibfnamefont {L.}~\bibnamefont
  {Hammer}}, \bibinfo {author} {\bibfnamefont {T.}~\bibnamefont {Kißlinger}},
  \bibinfo {author} {\bibfnamefont {M.}~\bibnamefont {Wagner}}, \bibinfo
  {author} {\bibfnamefont {R.~B.}\ \bibnamefont {Neder}}, \bibinfo {author}
  {\bibfnamefont {M.}~\bibnamefont {Schmid}}, \bibinfo {author} {\bibfnamefont
  {U.}~\bibnamefont {Diebold}},\ and\ \bibinfo {author} {\bibfnamefont {M.~A.}\
  \bibnamefont {Schneider}},\ }\href@noop {} {\bibfield  {journal} {\bibinfo
  {journal} {Phys. Rev. B}\ }}\bibinfo {note} {In preparation.}\BibitemShut
  {Stop}%
\bibitem [{\citenamefont {Tersoff}\ and\ \citenamefont
  {Hamann}(1985)}]{Tersoff85}%
  \BibitemOpen
  \bibfield  {author} {\bibinfo {author} {\bibfnamefont {J.}~\bibnamefont
  {Tersoff}}\ and\ \bibinfo {author} {\bibfnamefont {D.~R.}\ \bibnamefont
  {Hamann}},\ }\href {https://doi.org/10.1103/PhysRevB.31.805} {\bibfield
  {journal} {\bibinfo  {journal} {Phys. Rev. B}\ }\textbf {\bibinfo {volume}
  {31}},\ \bibinfo {pages} {805} (\bibinfo {year} {1985})}\BibitemShut
  {NoStop}%
\bibitem [{\citenamefont {Henkelman}\ \emph {et~al.}(2000)\citenamefont
  {Henkelman}, \citenamefont {Uberuaga},\ and\ \citenamefont
  {Jónsson}}]{Henkelman2000}%
  \BibitemOpen
  \bibfield  {author} {\bibinfo {author} {\bibfnamefont {G.}~\bibnamefont
  {Henkelman}}, \bibinfo {author} {\bibfnamefont {B.~P.}\ \bibnamefont
  {Uberuaga}},\ and\ \bibinfo {author} {\bibfnamefont {H.}~\bibnamefont
  {Jónsson}},\ }\href {https://doi.org/10.1063/1.1329672} {\bibfield
  {journal} {\bibinfo  {journal} {J. Chem. Phys.}\ }\textbf {\bibinfo {volume}
  {113}},\ \bibinfo {pages} {9901} (\bibinfo {year} {2000})}\BibitemShut
  {NoStop}%
\bibitem [{Sup()}]{SupMat}%
  \BibitemOpen
  \href@noop {} {}\bibinfo {note} {See Supplemental Material at [URL] for the
  LEED best-fit atomic coordinates in POSCAR format.}\BibitemShut {Stop}%
\bibitem [{\citenamefont {Kraushofer}\ \emph {et~al.}(2025)\citenamefont
  {Kraushofer}, \citenamefont {Imre}, \citenamefont {Franceschi}, \citenamefont
  {Ki\ss{}linger}, \citenamefont {Rheinfrank}, \citenamefont {Schmid},
  \citenamefont {Diebold}, \citenamefont {Hammer},\ and\ \citenamefont
  {Riva}}]{ViPErLEED_I}%
  \BibitemOpen
  \bibfield  {author} {\bibinfo {author} {\bibfnamefont {F.}~\bibnamefont
  {Kraushofer}}, \bibinfo {author} {\bibfnamefont {A.~M.}\ \bibnamefont
  {Imre}}, \bibinfo {author} {\bibfnamefont {G.}~\bibnamefont {Franceschi}},
  \bibinfo {author} {\bibfnamefont {T.}~\bibnamefont {Ki\ss{}linger}}, \bibinfo
  {author} {\bibfnamefont {E.}~\bibnamefont {Rheinfrank}}, \bibinfo {author}
  {\bibfnamefont {M.}~\bibnamefont {Schmid}}, \bibinfo {author} {\bibfnamefont
  {U.}~\bibnamefont {Diebold}}, \bibinfo {author} {\bibfnamefont
  {L.}~\bibnamefont {Hammer}},\ and\ \bibinfo {author} {\bibfnamefont
  {M.}~\bibnamefont {Riva}},\ }\href
  {https://doi.org/10.1103/PhysRevResearch.7.013005} {\bibfield  {journal}
  {\bibinfo  {journal} {Phys. Rev. Res.}\ }\textbf {\bibinfo {volume} {7}},\
  \bibinfo {pages} {013005} (\bibinfo {year} {2025})}\BibitemShut {NoStop}%
\bibitem [{\citenamefont {Rundgren}(2003)}]{Rundgren2003}%
  \BibitemOpen
  \bibfield  {author} {\bibinfo {author} {\bibfnamefont {J.}~\bibnamefont
  {Rundgren}},\ }\href {https://doi.org/10.1103/PhysRevB.68.125405} {\bibfield
  {journal} {\bibinfo  {journal} {Phys. Rev. B}\ }\textbf {\bibinfo {volume}
  {68}},\ \bibinfo {pages} {125405} (\bibinfo {year} {2003})}\BibitemShut
  {NoStop}%
\bibitem [{\citenamefont {Blum}\ and\ \citenamefont {Heinz}(2001)}]{Blum2001}%
  \BibitemOpen
  \bibfield  {author} {\bibinfo {author} {\bibfnamefont {V.}~\bibnamefont
  {Blum}}\ and\ \bibinfo {author} {\bibfnamefont {K.}~\bibnamefont {Heinz}},\
  }\href {https://doi.org/http://dx.doi.org/10.1016/S0010-4655(00)00209-5}
  {\bibfield  {journal} {\bibinfo  {journal} {Computer Physics Communications}\
  }\textbf {\bibinfo {volume} {134}},\ \bibinfo {pages} {392} (\bibinfo {year}
  {2001})}\BibitemShut {NoStop}%
\bibitem [{\citenamefont {Pendry}(1980)}]{Pendry1980}%
  \BibitemOpen
  \bibfield  {author} {\bibinfo {author} {\bibfnamefont {J.~B.}\ \bibnamefont
  {Pendry}},\ }\href {https://doi.org/10.1088/0022-3719/13/5/024} {\bibfield
  {journal} {\bibinfo  {journal} {J. Phys. C}\ }\textbf {\bibinfo {volume}
  {13}},\ \bibinfo {pages} {937} (\bibinfo {year} {1980})}\BibitemShut
  {NoStop}%
\bibitem [{\citenamefont {Kresse}\ and\ \citenamefont
  {Furthm\"uller}(1996)}]{vasp3}%
  \BibitemOpen
  \bibfield  {author} {\bibinfo {author} {\bibfnamefont {G.}~\bibnamefont
  {Kresse}}\ and\ \bibinfo {author} {\bibfnamefont {J.}~\bibnamefont
  {Furthm\"uller}},\ }\href {https://doi.org/10.1103/PhysRevB.54.11169}
  {\bibfield  {journal} {\bibinfo  {journal} {Phys. Rev. B}\ }\textbf {\bibinfo
  {volume} {54}},\ \bibinfo {pages} {11169} (\bibinfo {year}
  {1996})}\BibitemShut {NoStop}%
\bibitem [{\citenamefont {Perdew}\ \emph {et~al.}(1996)\citenamefont {Perdew},
  \citenamefont {Burke},\ and\ \citenamefont {Ernzerhof}}]{PBE}%
  \BibitemOpen
  \bibfield  {author} {\bibinfo {author} {\bibfnamefont {J.~P.}\ \bibnamefont
  {Perdew}}, \bibinfo {author} {\bibfnamefont {K.}~\bibnamefont {Burke}},\ and\
  \bibinfo {author} {\bibfnamefont {M.}~\bibnamefont {Ernzerhof}},\ }\href
  {https://doi.org/10.1103/PhysRevLett.77.3865} {\bibfield  {journal} {\bibinfo
   {journal} {Phys. Rev. Lett.}\ }\textbf {\bibinfo {volume} {77}},\ \bibinfo
  {pages} {3865} (\bibinfo {year} {1996})}\BibitemShut {NoStop}%
\bibitem [{\citenamefont {Perdew}\ \emph {et~al.}(2008)\citenamefont {Perdew},
  \citenamefont {Ruzsinszky}, \citenamefont {Csonka}, \citenamefont {Vydrov},
  \citenamefont {Scuseria}, \citenamefont {Constantin}, \citenamefont {Zhou},\
  and\ \citenamefont {Burke}}]{PBEsol}%
  \BibitemOpen
  \bibfield  {author} {\bibinfo {author} {\bibfnamefont {J.~P.}\ \bibnamefont
  {Perdew}}, \bibinfo {author} {\bibfnamefont {A.}~\bibnamefont {Ruzsinszky}},
  \bibinfo {author} {\bibfnamefont {G.~I.}\ \bibnamefont {Csonka}}, \bibinfo
  {author} {\bibfnamefont {O.~A.}\ \bibnamefont {Vydrov}}, \bibinfo {author}
  {\bibfnamefont {G.~E.}\ \bibnamefont {Scuseria}}, \bibinfo {author}
  {\bibfnamefont {L.~A.}\ \bibnamefont {Constantin}}, \bibinfo {author}
  {\bibfnamefont {X.}~\bibnamefont {Zhou}},\ and\ \bibinfo {author}
  {\bibfnamefont {K.}~\bibnamefont {Burke}},\ }\href
  {https://doi.org/10.1103/PhysRevLett.100.136406} {\bibfield  {journal}
  {\bibinfo  {journal} {Phys. Rev. Lett.}\ }\textbf {\bibinfo {volume} {100}},\
  \bibinfo {pages} {136406} (\bibinfo {year} {2008})}\BibitemShut {NoStop}%
\bibitem [{\citenamefont {Klime\ifmmode~\check{s}\else \v{s}\fi{}}\ \emph
  {et~al.}(2011)\citenamefont {Klime\ifmmode~\check{s}\else \v{s}\fi{}},
  \citenamefont {Bowler},\ and\ \citenamefont {Michaelides}}]{optB86b-vdW}%
  \BibitemOpen
  \bibfield  {author} {\bibinfo {author} {\bibfnamefont {J.~c.~v.}\
  \bibnamefont {Klime\ifmmode~\check{s}\else \v{s}\fi{}}}, \bibinfo {author}
  {\bibfnamefont {D.~R.}\ \bibnamefont {Bowler}},\ and\ \bibinfo {author}
  {\bibfnamefont {A.}~\bibnamefont {Michaelides}},\ }\href
  {https://doi.org/10.1103/PhysRevB.83.195131} {\bibfield  {journal} {\bibinfo
  {journal} {Phys. Rev. B}\ }\textbf {\bibinfo {volume} {83}},\ \bibinfo
  {pages} {195131} (\bibinfo {year} {2011})}\BibitemShut {NoStop}%
\bibitem [{\citenamefont {HenkelmanGroup}(2020)}]{VTSTtools}%
  \BibitemOpen
  \bibfield  {author} {\bibinfo {author} {\bibnamefont {HenkelmanGroup}},\
  }\href@noop {} {\bibinfo {title} {Transition state tools for \textsc{VASP}}}
  (\bibinfo {year} {2020}),\ \bibinfo {note}
  {http://henkelmanlab.org/vtsttools/}\BibitemShut {NoStop}%
\bibitem [{\citenamefont {Ganduglia-Pirovano}\ and\ \citenamefont
  {Scheffler}(1999)}]{Scheffler1999}%
  \BibitemOpen
  \bibfield  {author} {\bibinfo {author} {\bibfnamefont {M.~V.}\ \bibnamefont
  {Ganduglia-Pirovano}}\ and\ \bibinfo {author} {\bibfnamefont
  {M.}~\bibnamefont {Scheffler}},\ }\href
  {https://doi.org/10.1103/PhysRevB.59.15533} {\bibfield  {journal} {\bibinfo
  {journal} {Phys. Rev. B}\ }\textbf {\bibinfo {volume} {59}},\ \bibinfo
  {pages} {15533} (\bibinfo {year} {1999})}\BibitemShut {NoStop}%
\end{thebibliography}%

\end{document}